# Inverse Spin Hall Effect Induced by Asymmetric Illumination of Light on Topological Insulator $Bi_2Se_3$

Di Fan[*], Rei Hobara, Ryota Akiyama[†], and Shuji Hasegawa

*Department of Physics, The University of Tokyo, Tokyo 113-0033, Japan*

**Abstract**
Using circularly polarized light is an alternative to electronic ways for spin injection into materials. Spins are injected at a point of the light illumination, and then diffuse and spread radially due to the in-plane gradient of the spin density. This diffusion is converted into a circular charge current by the inverse spin Hall effect (ISHE). With shining the circularly polarized light at asymmetric parts of the sample, such as near edges, we detected this current as a helicity-dependent component in the photocurrent. We present a model for this ISHE based on the experimental results and the finite-element-method (FEM) simulation of the potential distribution induced by spin injection. Our model shows that the ISHE photocurrent generates an electric dipole at the edge of the sample, causing the measured charge current. The asymmetric light-illumination shown here is a simple way to inject and manipulate spins, opening up a door for novel spintronic devices.

[*] fan@surface.phys.s.u-tokyo.ac.jp
[†] akiyama@surface.phys.s.u-tokyo.ac.jp

Spintronics is a promising approach to realize next-generation electronics devices in which spin injection and detection are the key factors. To convert spin into charge current, we can rely on the inverse spin Hall effect (ISHE) [1, 2] or the spin-momentum locking effect on the Fermi surface [3, 4]. Compared with the electrical spin injection, optical methods are non-invasive and contactless, which has advantages for remote control or inspection. Many attempts have been tried to use the laser to control electron spins [5, 6]. The circular photogalvanic effect (CPGE), which is an optical way to convert spin into a charge current, has been demonstrated in various materials [7-11].

On one hand, the emergence of the photo-induced inverse spin Hall effect (PISHE) can be date back to as early as 1980's [12]. They observed charge current generated by the ISHE with an external magnetic field to align the spin of photo-excited carriers to the desirable direction with respect to the out-of-plane gradient of the concentration of excited carriers. Recently, some groups report that the PISHE is realized in the two-dimensional electron gas (2DEG) with strong spin-orbit coupling [13] and the topological insulator [14] without magnetic field.

When spin is injected into a material, the diffusing spin current $J_d$ would be generated following the diffusion equation,

$$J_d = D \cdot \nabla N. \tag{1}$$

Here, $D$ and $N$ are the diffusion constant and the spin density, respectively. If the ISHE occurs in this situation, the induced charge current $J_{ISHE}$ can be written as,

$$J_{ISHE} = \theta_k J_d \times S_z, \tag{2}$$

where $\theta_k$ and $S_z$ are the spin Hall angle and the projection of the spin in $z$ direction (surface normal), respectively. When the spins are injected at a point by light illumination, the injected spins diffuse and spread out radially on the sample surface. Such a radial diffusion of the injected spins will, according to Eq. (2), yield a circular charge current when the spins are in $z$ direction. However, the electric current observed between the electrodes at both ends of the sample is zero because of "canceling out", when the light is irradiated at the center of the sample as shown in Fig. 1(a).

On the other hand, when the laser light illuminates the interface between two different materials, where the carrier diffusion in Material 1 is smaller than that in Material 2, as shown in Fig. 2(b), the charge currents induced by the ISHE are not equivalent in the two parts, and a net charge current flowing one direction would appear. For the simplest and most practical case, we can assume that Material 1 is vacuum and Material 2 is the sample to be investigated. Obviously, there is no diffusion current in the vacuum side, thus the circular electric current is expected to be detected by the electrodes at both ends of the sample as shown in Fig. 1(c).

In the present study, we succeeded in detecting this ISHE component rather than the CPGE component of the photocurrent, which is induced by illuminating circularly polarized light under normal incidence near the edge of the sample of a topological insulator $Bi_2Se_3$ thin film. Then, such ISHE component disappeared when the light shines around the center of the sample.

We chose a well-studied topological insulator $Bi_2Se_3$ because it possesses strong spin-orbit coupling for promising the ISHE. Previously, the CPGE was observed on $Bi_2Se_3$ flakes using lasers with the wavelength $\lambda$ of 780 nm and 1064 nm [15, 16], which showed the conversion of the in-plane spin to the charge current. In our experiments, the laser with $\lambda$ = 1550 nm was irradiated on a $Bi_2Se_3$ thin film of 13 quintuple-layer (QL) thick grown by molecular beam epitaxy (MBE) technique on Si(111) substrate [17]. The sample was capped with $Al_2O_3$ before transferred into the measurement chamber. More information about the sample can be found in the supplemental material (SM). Since the photon energy of 0.8 eV (1550 nm) is smaller than the band gap energy of the Si substrate, the observed photocurrent is originated from the $Bi_2Se_3$ film only. During the experiment, the laser power was kept at 1.75 mW which was confirmed to make no damage to the film. All measurements with laser illumination were performed in the MBE ultra-high-vacuum chamber at room temperature.

To check the spin direction, we changed the incident angle of the laser light with respect to the surface-normal, and found out that the incident angle dependence of the helicity-dependent photocurrent (HDP) was similar to that in the previous report [8] (see the SM). However, the HDP was non-zero even at normal incidence, especially when the light spot was near the edge of the sample. This cannot be explained by the CPGE mechanism reported in previous researches [8, 18, 19].

In this study, we used the circularly polarized light at normal incidence to inject the out-of-plane component of spin. The photocurrent $J$ detected with the electrode at both ends is related with the polarization of the light, and thus with the rotation angle $\phi$ of the quarter-wave plate (QWP) in a way,

$$J = C \sin 2\phi + L_1 \sin 4\phi + L_2 \cos 4\phi + D. \qquad (3)$$

Here, $\phi$ is the rotation angle of QWP, enabling to change the light polarization. The parameter $C$ is the HDP component related with the circularly polarized light. In our experimental scheme, the circularly polarized light excites electrons and holes having the out-of-plane spin. The electrons and holes diffuse in the same directions along each concentration gradient. The diffusion of carriers with spins is converted into a charge current by the ISHE. Therefore the parameter $C$ represents the ISHE component only, because at normal incidence of light the CPGE should not contribute to the HDP due to the $D_{3d}$ crystal symmetry of $Bi_2Se_3$ (see the SM).

The parameters $L_1$ and $L_2$ indicate the photocurrent generated by the linearly polarized light. Specifically, $L_1$ represents the linear photogalvanic effect (LPGE), which arises in a crystal without inversion center and it is related with the crystalline symmetry [19]. On the other hand, $L_2$ is out of the expectation of the LPGE and some reports explain it as the anomalous linear photogalvanic effect (ALPGE) [20]. $D$ represents an off-set current which is not related with polarization of the light and it may originate from the photo Dember effect or the thermal effect.

Figure 2 shows the QWP-angle dependences of the photocurrent detected by the electrodes at both ends of the sample, where the laser light illuminates the right edge, the center, and the left edge of the sample, respectively. The amplitudes of the photocurrent at the left and right edges are significantly larger than that at the center. Moreover, the HDP (the difference between the right- and left-handed circular polarizations indicated by the difference between orange and blue dashed lines) is opposite between the left- and right-edge irradiations whereas it is negligible for the center irradiation.

The dependence of the fitting parameter $C$ in Eq. (3) on the position of laser illumination across the sample surface is shown in Fig. 3. When the laser spot (ca. 1.2 mm in diameter) was focused on the sample surface, the charge current induced by the ISHE was enhanced at both edges of the sample, while it was almost zero at the center as indicated by red circles. The sign of parameter $C$ is opposite on the opposite sides of the sample. This is consistent with the expectation of Fig. 1(c). Then, when the laser was defocused (laser spot of ca. 2.4 mm in diameter), as shown by blue triangles in Fig. 3, it showed much smaller values and smoother change in $C$ parameter.

Figure 4 shows the results of the finite-element method (FEM) simulation for visualizing the potential distribution on the sample which is induced by spin injection (a) at the edge or (b) at the center on the sample. Equations used for the calculation of ISHE are described as Eqs. (10) - (13) in Ref. [21]. As shown in Fig. 4(b), when the spin injection takes place by illuminating laser at the center of the sample, no electric potential is generated. On the other hand, as in Fig. 4 (a), when the spins are injected on the left edge of the sample, remarkably, a finite electric potential due to an electric dipole is created by the ISHE. The dipole center is the spin injection point, corresponding to the spot center of light, and the two peaks of the positive and negative electric potential are at the boundaries of the spin injection area, corresponding to the laser spot area. This potential distribution can be interpreted as that the circular charge current path induced by the ISHE acts like a conveyer; carriers are transferred from one part of the edge to the other along the half-circular trajectory. Consequently, an electric potential builds up due to the carrier accumulation. When the field of this electric dipole reaches to balance the ISHE under continuous illumination of laser, no more carrier are accumulated and the system is in equilibrium, resulting in a potential difference between two electrodes at both ends of the sample.

During the optical excitation, electron-hole pairs are generated by the circular polarized light and these carriers diffuse outward from the spot center. They have the same spin which are transferred from the angular momentum of the circularly polarized light. Usually the diffusion coefficient of electrons is larger than that of holes, thus electrons diffuse down the gradient faster, leading to a density imbalance between electrons and holes, which creates an electric field. This electric field acts to decelerate electrons and accelerate holes until both diffusions reach the same rate. In this equilibrium, there is no net charge current along the carrier diffusion direction (so-called ambipolar diffusion [22]). If the carrier lifetime is long enough to fulfill this equilibrium, a pure spin current is created by photoexcitation. The electrons and holes with the same spin will flow oppositely along the half-circular ISHE path, leading to enhancement of the electric dipole field compared with the case of electrons only.

The accumulation of carriers near edges should strongly depend on the laser spot profile and the distance between the spot center and the sample edge. In our system, the spin diffusion length is much smaller than the spot size. Hence, the region where the spin current flows is determined by the spot size. Consequently, the dipole distance is restrained by the spot size and the dipole appears only when the circular spot is partially cut by the edge of the sample. To see how the laser spot size and its position on the sample surface affect the photocurrent, a simulation result is shown in Fig. 5. The position dependence indicates that the larger peak value is acquired by the more focused (smaller) laser spot. This is consistent with the experimental results in Fig. 3.

In conclusion, we observed the ISHE induced by the circularly polarized infrared light at normal incidence on $Bi_2Se_3$ thin film. The behavior of the photocurrent by the ISHE can be explained by using the FEM simulation. It suggests that an electric dipole is generated at the edge of the sample, resulting from the accumulation of carriers steered by the ISHE. Therefore, the current we measured came from the potential of this electric dipole. The ISHE induced by asymmetric manner of light illumination shown in this study will pave a way for novel spintronics devices operating at room temperature without external magnetic field, and be helpful for an easier accessible spin-charge conversion.

## Acknowledgement

We acknowledge financial supports from the Japan Society of Promotion of Science (KAKENHI 16H02108, 25110010, 25246025, 22246006, JP16H00983, JP15K21717) and MEXT (Grant-in-Aid for Scientific Research on Innovative Areas "Molecular Architectonics" 25110001). DF also likes to thank for the financial support from The Advanced Leading Graduate Course for Photon Science (ALPS) of the MEXT "Program for Leading Graduate Schools".

# Figures

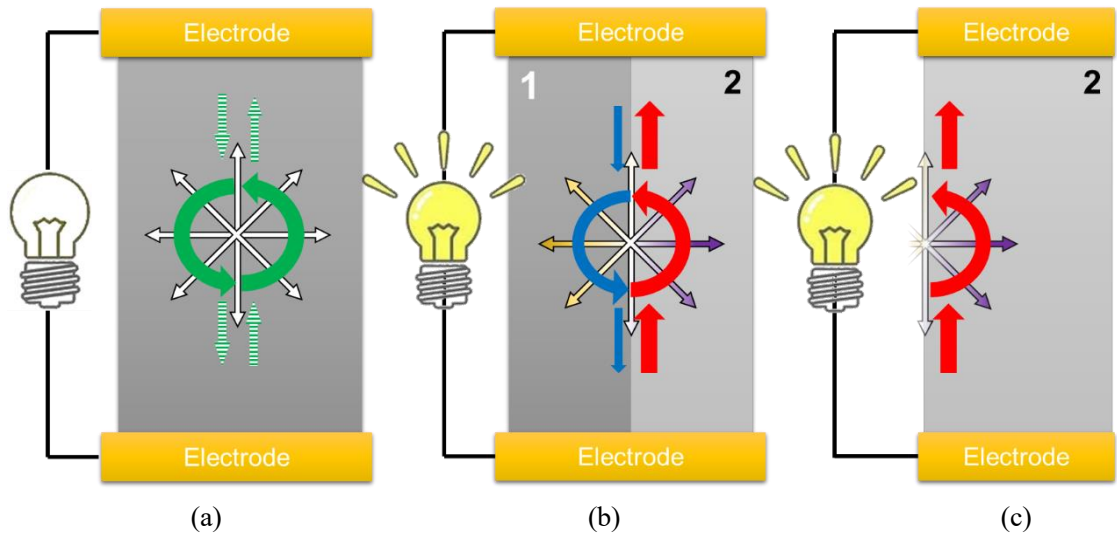

(a) (b) (c)

FIG. 1 (a) When the circularly polarized light irradiates the center on the sample surface, the circular charge current (green solid arrows), which is converted from the radial spin current (white solid arrows) by the ISHE, appears. However, no outcome can be detected by the electrodes at both ends of the sample. The dashed green arrows are the inflow and outflow of the closed circular electric current, and the net flow along the sample is zero. (b) When the circularly polarized light irradiates an interface between different materials, 1 and 2, the inflow and outflow currents are not equivalent, resulting in a net current which flows along the longitudinal direction. (c) As one of the simplest situations of (b), when the material 1 is vacuum, the net current is expected to be the maximum.

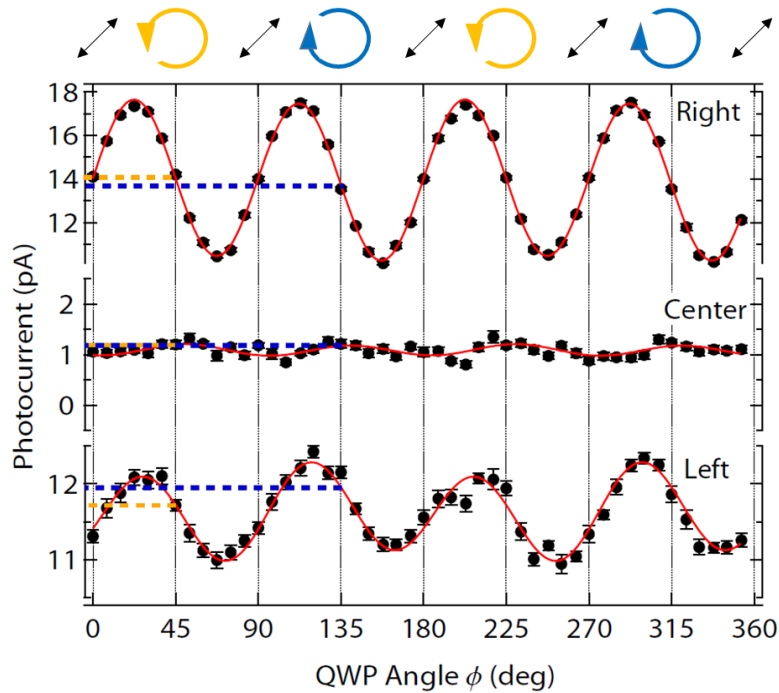

FIG. 2 Polarization dependences of the photocurrent measured at normal incidence of light. Polarization was changed by rotating the QWP. The points were the average of 20 times scan raw data. Red lines are fitting curves by Eq. (3). The orange (blue) dashed line indicates the photocurrent at the right-handed (left-handed) circular polarization. The difference between the two dashed lines corresponds to two times of the magnitude of parameter $C$.

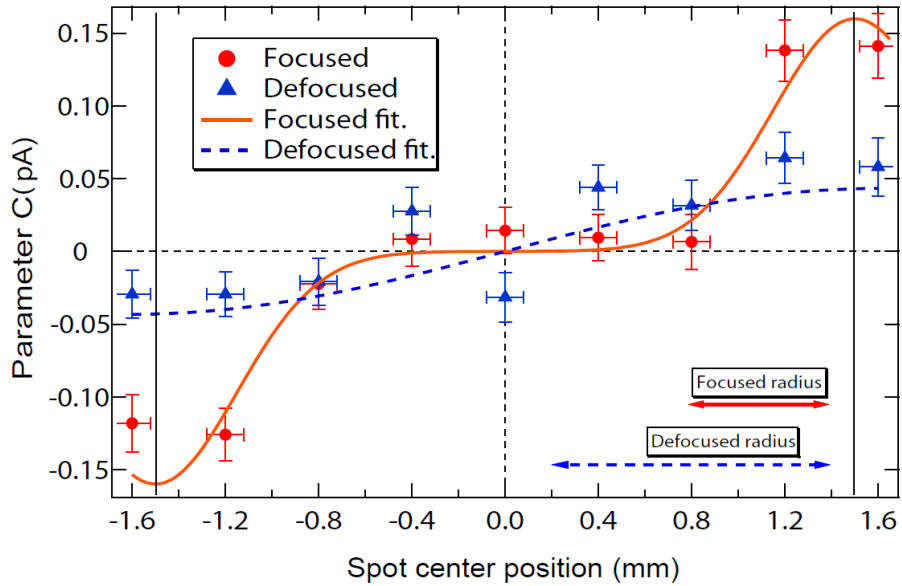

FIG. 3 Dependence of the parameter $C$ on the spot position across the sample, with the focused (red circles) and defocused (blue triangles) laser at normal incidence. The sample edges are shown as the vertical lines at $\pm 1.5$ mm.. The radius with the focused and defocused laser spot is estimated to be ~ 0.6 mm and ~ 1.2 mm, respectively, by fittings. The curves are simulation results shown in Fig. 5.

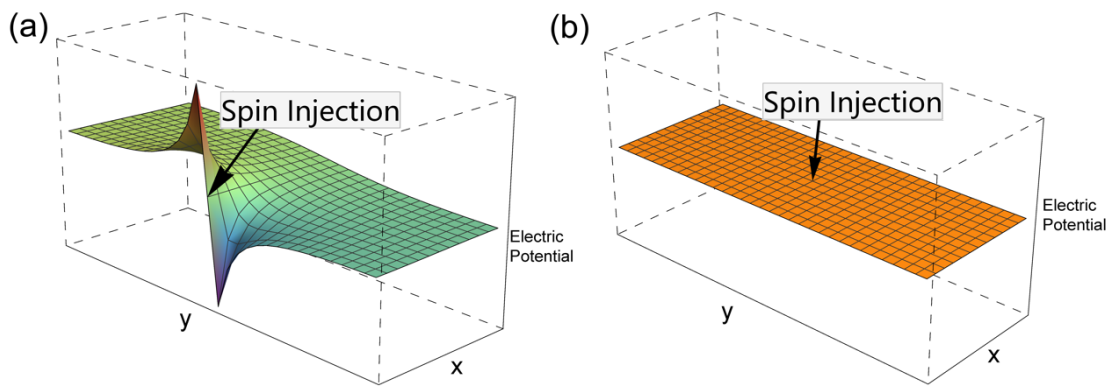

FIG. 4 FEM simulations of the electrical potential distribution at equilibrium. The $x$- and $y$-axes are of the sample scale, whereas the electric potential is in arbitrary unit. The arrows point spin injection positions. (a). A situation where the spin injection occurs at the left edge of the sample. Two peaks are seen with opposite signs, which represent generation of an electric dipole at the left edge. (b). A situation where spins are injected at the center of the sample as indicated by the solid arrow. No potential

difference is generated.

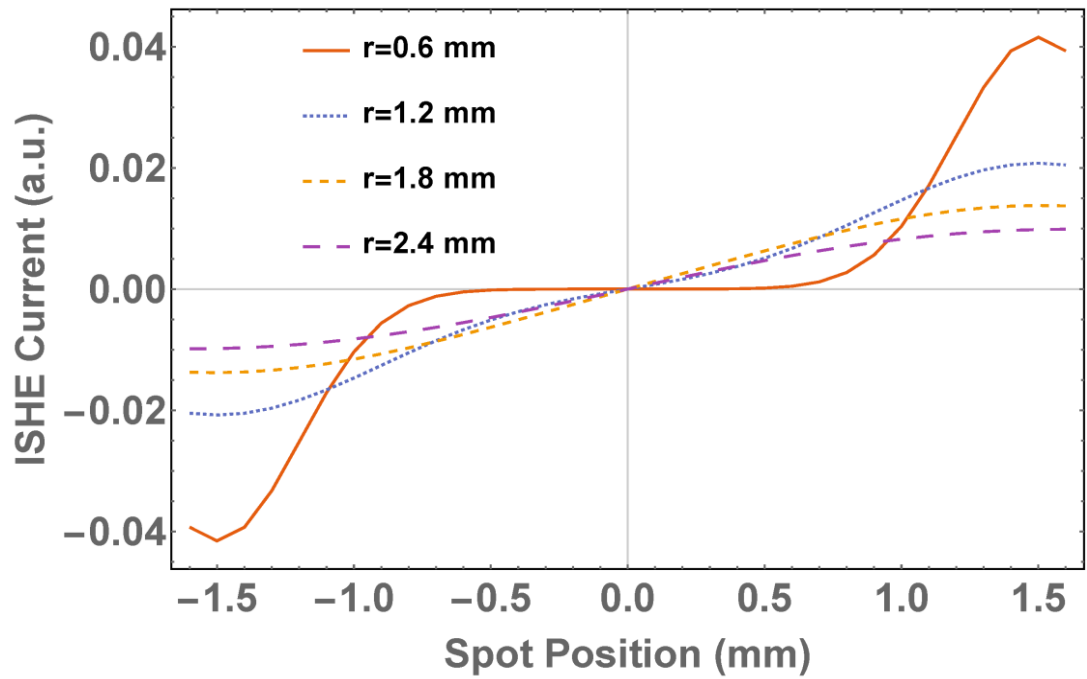

FIG. 5 Simulation results of the dependence of the ISHE current on the laser spot size and spot position across the sample surface. $r$ represents the radius ($1/e^2$ definition) of the laser spot in mm. When the spot size decreases, the ISHE current is enhanced at edges. The sample edge is at $\pm\ 1.5$ mm.

## Supplemental material

## Inverse Spin Hall Effect by Asymmetric Illumination of Light on Topological Insulator Bi$_2$Se$_3$

Di Fan[‡], Rei Hobara, Ryota Akiyama[§], and Shuji Hasegawa

*Department of Physics, The University of Tokyo, Tokyo 113-0033, Japan*

**Sample preparation**

The thin film of Bi$_2$Se$_3$ was grown on a Si(111) substrate in a ultrahigh vacuum (UHV) chamber by the molecular beam epitaxy technique. First we prepared the Si(111)-β-√3 × √3-Bi surface structure by depositing one-atomic layer of Bi, and then proceeded to Bi$_2$Se$_3$ growth. The flux ratio of Bi to Se was between 1:10 to 1:20. The substrate was kept at ~ 175 °C during the growth of the first QL Bi$_2$Se$_3$, and after that the substrate temperature was raised and kept at ~ 200 °C for further growth. The growth rate was ~ 4 min/QL. The thickness was monitored by the RHEED oscillation *in-situ* as shown in Fig. S1. It is known that *n*-type Bi$_2$Se$_3$ is usually formed under such growth procedure due to Se vacancies.

From the previous angle-resolved photoemission spectroscopy (ARPES) reports, in addition to Dirac-cone type topological surface states, Rashba-type spin-split conduction bands of the film also exist at the Fermi surface of Bi$_2$Se$_3$ [1]. The hexagonal warping at the Rashba-type states and topological surface states [1, 2] in the vicinity of the Fermi surface play an important role for the out-of-plane spin component.

For *ex-situ* measurement, ~ 2ML Al was deposited on the Bi$_2$Se$_3$ thin film. Then, the sample was taken out of the UHV chamber, resulting in immediate formation of Al$_2$O$_3$ capping layer by oxidation [3].

**Circular photogalvanic effect (CPGE)**

The circular photogalvanic effect is related with the crystal symmetry of the sample. The symmetry group of the Bi$_2$Se$_3$ crystal is $D_{3d}$, which includes the inversion symmetry. Generally, photogalvanic effect occurs only in the crystal structure without inversion symmetry. Therefore, for Bi$_2$Se$_3$, the PGE is allowed only at the surface of Bi$_2$Se$_3$ where the inversion symmetry in *z*-direction is broken and the symmetry for the surface is $C_{3v}$. The second rank pseudo-tensor for $C_{3v}$ symmetry is

$$\begin{pmatrix} 0 & \gamma_{xy} \\ -\gamma_{xy} & 0 \end{pmatrix}. \tag{S1}$$

The CPGE current along *y*-direction on the surface is expected when the incident plane of the light is set in *xz*-plane. It is written as [4]:

$$j_y^{CPGE} = -\gamma_{xy} \hat{e}_x P_{circ} E_0^2, \tag{S2}$$

where $P_{circ}$ is the degree of circular polarization, $E_0$ is the electric field amplitude, and $\hat{e}_x$ is the *x*

---

[‡] fan@surface.phys.s.u-tokyo.ac.jp
[§] akiyama@surface.phys.s.u-tokyo.ac.jp

projection of the unit vector pointing in the direction of the light propagation ($\hat{e}_x = \sin\theta$, where $\theta$ is the angle between the light direction and the surface-normal direction). Then, the CPGE is expected to disappear under normal incidence, *i.e.* incident along *z*-direction, because $\hat{e}_x$ is zero.

For a 13QL-Bi$_2$Se$_3$ thin film, both the incident-angle dependence and the irradiation-position dependence were measured as shown in Fig. S2. The parameter *C* for the incidence angle of $\pm$ 40° along *x*-direction are opposite to each other in sign, which is exactly the nature of the CPGE; as shown in Eq. (S2), the CPGE photocurrent reverses its sign when the incident direction (in-plane component $\hat{e}_x$) is reversed. On the other hand, the clear position dependence infers a different mechanism other than the CPGE occurring. As shown in Fig. S2, the tendency of the position dependence does not change despite the incident angle changes, which suggests that the position dependence is not related with the in-plane component of the angular momentum of the light.

## Symmetry breaking ISHE induced by the circularly polarized light

As is known, the laser irradiance distribution in the spot is a Gaussian. Since a Gaussian wave propagates along *z*-axis, the irradiance at a distance *r* in *xy* plane from the center of spot is expressed as:

$$I(r, z_0) = I_0 \exp(\frac{-r^2}{2\sigma^2}), \tag{S3}$$

where $2\sigma$ indicates the radius of the laser spot (1/e$^2$ definition). The spot center is at ($x_0$, $y_0$) on *xy* plane, and the incident angle $\theta$ with respect to the surface-normal direction is in *xz*-plane, the intensity distribution *I* is given as

$$I = I_0 G(x, y, \theta), \tag{S4}$$

$$G(x, y, \theta) = \frac{\cos\theta}{2\sigma^2\pi} \exp(-\frac{(x-x_o)^2 \cos^2\theta + (y-y_0)^2}{2\sigma^2}). \tag{S5}$$

This directly influences the distribution of the photoexcited carriers. Moreover, the carrier distribution generates a diffusing spin current $\boldsymbol{J}_d$ along the concentration gradient. According to the diffusion equation,

$$\boldsymbol{J}_d = D \cdot \nabla N. \tag{S6}$$

Here *N* is the concentration of photoexcited carriers which, as stated above, should be proportional to the irradiance of the light. Therefore, it is represented by the same Gaussian distribution $G(x, y, \theta)$. *D* is the diffusion coefficient for the ambipolar diffusion of photoexcited carriers. For simplicity, the spot center is at ($x_0$, 0) on *x*-axis. Then we can substitute the $G(x, y, \theta)$ in Eq. (S6) to obtain the diffusion spin current at (*x*, *y*) in a vector form and $\boldsymbol{J}_d$ is given as

$$\boldsymbol{J}_d = -AD \left\{ \frac{(x-x_o)\cos^3\theta}{2\sigma^4\pi} \exp\left(-\frac{(x-x_o)^2 \cos^2\theta + y^2}{2\sigma^2}\right), \frac{y\cos\theta}{2\sigma^4\pi} \exp\left(-\frac{(x-x_o)^2 + y^2}{2\sigma^2}\right) \right\}. \tag{S7}$$

Here, *A* is a coefficient relating the light intensity with the number of the photoexcited carrier. Eq. (S7) means that the spin current flows out radially from the spot center.

Due to the ISHE, this spin current generates charge current $J_{ISHE}$ in a way,

$$J_{ISHE} = \theta_k \cdot \boldsymbol{J}_d \times \boldsymbol{S_z}, \tag{S8}$$

where $\theta_k$ is the spin Hall angle and $\boldsymbol{S_z}$ is the *z*-component of spin of the photoexcited carriers. The direction of the charge current is always perpendicular to the direction of the spin current. Then, since the spin current flows radially from the laser spot center, the charge current flows circularly around the laser spot. In Fig. S3, the diffusion spin current (red arrows) and charge current induced by the ISHE (yellow arrows) are plotted together with irradiance of the laser (red circle). To get rid of the CPGE and to focus on the ISHE only, we set $\theta$ to be zero, which is the case of normal incidence.

As we suggested in the main article, the system goes to equilibrium and there is no charge transfer from one part to another at the edge; the ISHE current at the edge is balanced by the electric field due to the dipole. From this point, Poisson equation for the whole sample gives the potential distribution $\varphi(x,y)$ when the laser spot center is at the left edge of the sample as

$$\nabla(\cdot \sigma_{1 \text{ or } 2} \nabla \varphi(x,y)) = 0, \tag{S9}$$

$$\sigma_2 \boldsymbol{n} \cdot \nabla \varphi(x,y)|_{x=-w} = -J_x^{ISHE}(-w,y), \tag{S10}$$

$$\varphi(x,0) = 0. \tag{S11}$$

Note that $\boldsymbol{n}$ is the unit vector perpendicular to the left edge of the sample, and $\sigma_1$ and $\sigma_2$ are the conductance of the metallic electrodes and sample, respectively. $w$ is the half of the sample width and thus the left edge is $x = -w$. Eq. (S11) is assumed by the symmetry of the system. From Eqs. (S9) - (S11), the numerical solution for the potential distribution on the sample surface was obtained as shown in Fig. S4. This result shows almost the same distribution as the FEM calculation in the main text.

It is difficult to obtain the analytical solution of the equations above, thus we tried another way for describing the electric potential of our system. Here, we assume that the charges exist only on the sample edge without electrodes (*i.e.* along *y*-direction in Fig. S3), and the charge density is proportional to the amplitude of the ISHE current. When the dipole distance is small enough the dipole electric potential can be expressed as,

$$\varphi(r) = \frac{kp\cos\beta}{r^2} \tag{S12}$$

where $k$ is Coulomb's constant, $\boldsymbol{p} = q\boldsymbol{d}$ is defined as the dipole moment ($\boldsymbol{d}$ is the displacement vector pointing from the negative charge $-q$ to the positive charge $+q$), $\beta$ is the angle between $\boldsymbol{r}$ and $\boldsymbol{p}$, and $r$ is the distance from the dipole center. As mentioned above, for the $J^{ISHE}$ at both side of the sample ($w, y$) or ($-w, y$), $\boldsymbol{p}$ should be

$$\boldsymbol{p} = J_{x=\pm w}^{ISHE} y dy, \tag{S13}$$

and then the voltage between the two electrodes at both ends of the sample is

$$V = 2k \int_0^l \frac{J_{x=w}^{ISHE} y dy \cos\theta}{r^2} + 2k \int_0^l \frac{J_{x=-w}^{ISHE} y dy \cos\theta}{r^2} . \tag{S14}$$

Here, $l$ is the half length of the sample along *y*-direction. The potential distribution along *x*-direction at the same $y$ is similar, assuming that the potential at equilibrium is just smoothed from the potential at the same $y$. Thus, the final expression for the voltage between the electrodes at both ends is

$$V = kAD \frac{e^{-\frac{2w+2x_0^2+l^2}{2\sigma^2}} \left(e^{\frac{(w-x_0)^2}{2\sigma^2}} - e^{\frac{(w+x_0)^2}{2\sigma^2}}\right)\left(2l - \sigma e^{\frac{l^2}{2\sigma^2}}\sqrt{2\pi}\text{Erf}\left(\frac{l}{\sigma\sqrt{2}}\right)\right)}{4\sigma^2 \pi n^2}. \tag{S15}$$

$(x_o, 0)$ is the spot position ($x_o \in [-w, w]$). By knowing the resistance of the sample $R$, then the measured current $J_y$ flowing between the electrodes is

$$J_y = \frac{V}{R}. \tag{S16}$$

From Eq. (S15), we are able to investigate the ISHE further. First, as we assumed earlier, the ISHE current should be linearly proportional to the light intensity as shown in Fig. S5. Moreover, we can use Eq. (S15) to fit the position dependence as in Fig. 3 of the main article and the spot size dependence, Fig. 5 in the main article, can also be obtained.

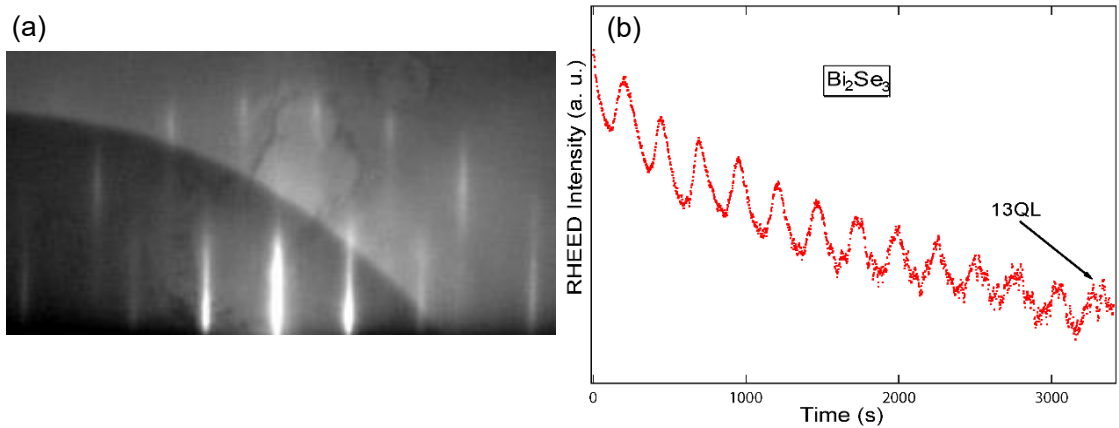

FIG. S1 (a) RHEED pattern taken after the growth of $Bi_2Se_3$. (b) Intensity oscillation of the specular spot in RHEED observed during $Bi_2Se_3$ growth.

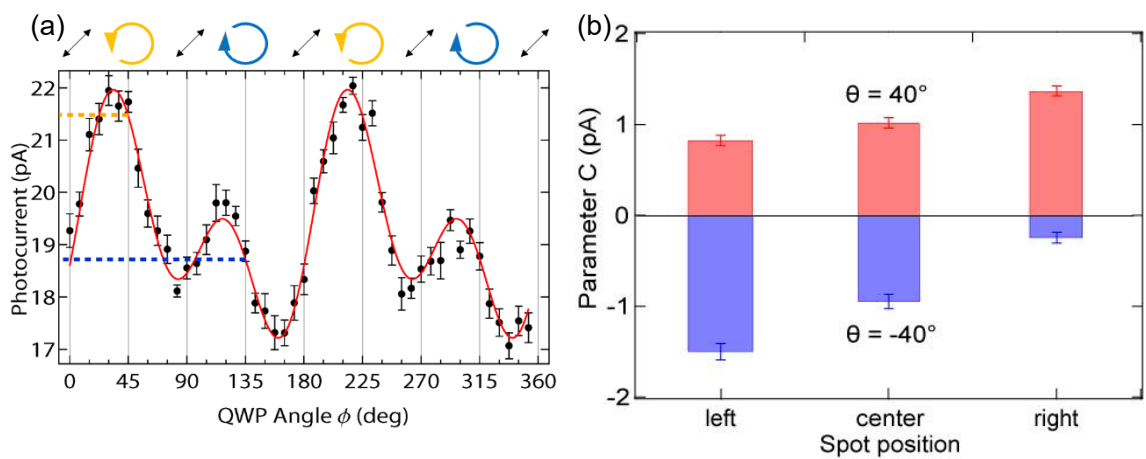

FIG. S2 (a) Polarization dependent photocurrent taken by irradiating laser light near the edge of the sample under incident angle $\theta = 40°$, and the curve is the fitting by Eq. (3). (b) Spot position (near left and right edges and center of the sample) dependence of the parameter $C$ estimated from the fitting at the incident angle $\theta$ of 40° and -40° with respect to the surface-normal direction. The sign of parameter C is reversed when the incident angle is reversed, but the position dependence is in the same trend for both incident angles.

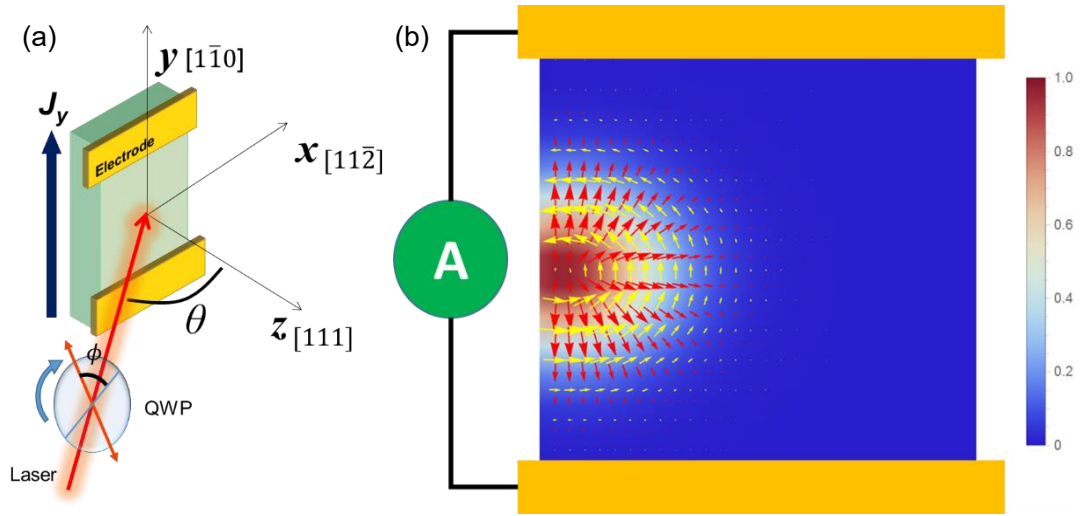

FIG. S3 (a) Schematic picture of the experimental setup. The coordinate based on a silicon substrate is shown; $x$-axis along [112], $y$-axis along [110], and $z$-axis along [111] directions, respectively. $xz$ plane is the plane of incidence of light. Incident angle $\theta$ is the angle between the laser propagation direction and $z$-axis. By rotating the quarter-wave plate (QWP), the polarization of the initially linearly polarized laser is changed. The photocurrent $J_y$ was measured along $y$-axis by the electrodes at both ends. (b) A simulation result of the ISHE by the finite element method. The color gradient shows the distribution of the laser intensity, and thus it indicates the distribution of the photo-excited carrier density (and also spin density). Red arrows show the diffusion direction of carriers and spin, and the yellow arrows indicate the charge current induced by the ISHE. Arrow lengths indicate the intensities of respective currents.

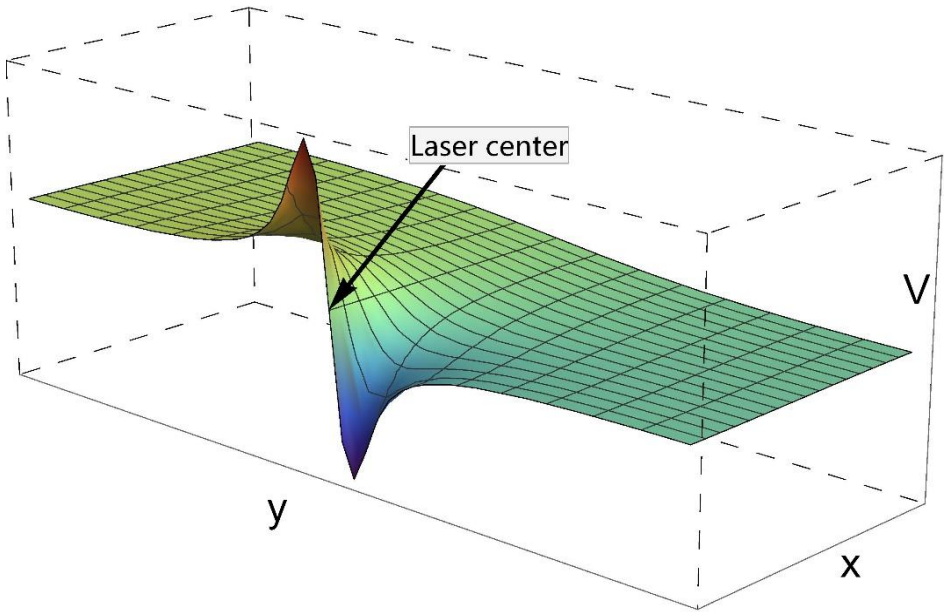

FIG. S4 By solving the Poisson equation Eq. (S9) with the boundary conditions Eqs. (S10) and (S11), the electric potential distribution can be calculated. The numerical solution is plotted as a color plot, and it is similar to the potential distribution calculated by the FEM.

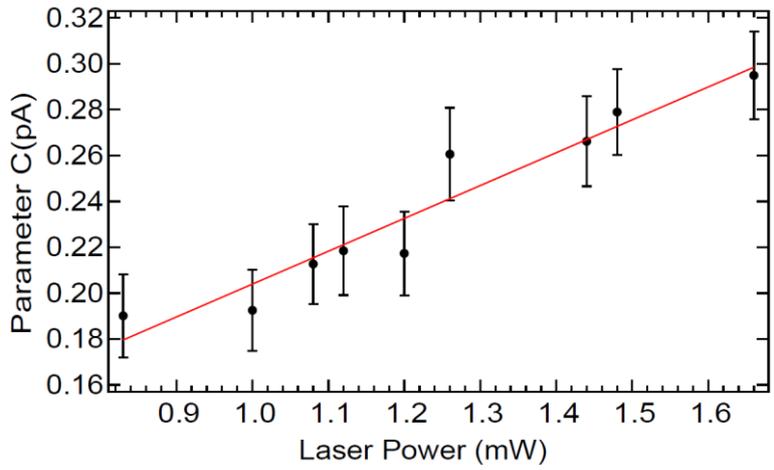

FIG. S5 Laser power dependence of the ISHE current (parameter C) at the normal incidence. The laser spot was focused to the minimum size and the spot position was near the sample edge to maximize the ISHE.